\documentclass[aps,twocolumn,floats,prb]{revtex4}
\usepackage{graphics,dcolumn}
\begin{document}

\title{Static and dynamical properties of heavy water at ambient conditions from
first-principles molecular dynamics}
\author{P. H-L. Sit}
\affiliation{Department of Physics, Massachusetts Institute of Technology, Cambridge MA 02139}
\author{Nicola Marzari}
\affiliation{Department of Materials Science and Engineering, Massachusetts Institute of Technology, Cambridge MA 02139}

\begin{abstract}
The static and dynamical properties of heavy water have been studied at ambient 
conditions with extensive Car-Parrinello molecular-dynamics simulations in the 
canonical ensemble, with temperatures ranging between 325 K and 400 K. 
Density-functional theory, paired with a modern exchange-correlation functional 
(PBE), provides an excellent agreement for the structural properties and binding 
energy of the water monomer and dimer. On the other hand, the structural and 
dynamical properties of the bulk liquid show a clear enhancement of the local 
structure compared to experimental results; a distinctive transition to liquid-like 
diffusion occurs in the simulations only at the elevated temperature of 400 K. 
Extensive runs of up to 50 picoseconds are needed to obtain well-converged thermal 
averages; the use of ultrasoft or norm-conserving pseudopotentials and the larger 
plane-wave sets associated with the latter choice had, as expected, only negligible 
effects on the final result. Finite-size effects in the liquid state are found to be 
mostly negligible for systems as small as 32 molecules per unit cell.
 
\end{abstract}

\date{\today}
\maketitle

\section{Introduction}

Water, due to its abundance on the planet and its role in many of the organic and inorganic 
chemical processes, has been studied extensively and for decades both at the theoretical and at the
experimental level \cite{water}$^-$\cite{CPwatend}.
The peculiar interplay of hydrogen bonding, glassy behavior, and of
quantum-mechanical effects on the dynamics of the atomic nuclei make computer
simulations challenging, and a great effort has been expended to
build a comprehensive and consistent microscopic picture, and a link with observed 
macroscopic properties \cite{emp1}$^-$\cite{CPwatend}.
Additionally, it is only recently that close agreement for 
fundamental structural information such as the radial distribution function
has been obtained between different experimental techniques, such as 
X-ray \cite{xray} and neutron diffraction \cite{neutron} measurements.

Computational studies based on molecular dynamics simulations have also a
rich history in the field. 
Simulations using force-fields models\cite{emp1}$^-$\cite{empend} have been successful at 
reproducing many structural and dynamical properties of liquid water.
However, empirical models rely on parameters which are determined by fits to known experimental data,
or occasionally to ab-initio results. Their transferability to different environments, or the ability to
reproduce faithfully the microscopic characteristics of hydrogen bonding, are often in question. 
Due to development of novel techniques \cite{CP1}$^,$\cite{CP2} and the ever-increasing improvement in computational power,
extensive molecular-dynamics simulations from first-principles are now possible. The increased accuracy and
predictive power of these simulations comes at a significant price, and careful considerations has
to be given to the length scales and time scales that can be afforded in a first-principle simulation, and
the trade-offs in statistical errors when compared with classical simulations.
Numerous ab-initio simulations on water have appeared \cite{CPwat1}$^-$\cite{CPwatend}, showing good agreement 
with experiments for the structural and dynamical data.
Recent results have also reported \cite{GrossmanI}$^,$\cite{kress} that careful equilibration for
ab-initio water at ambient conditions leads to radial distribution functions over-structured in
comparison with experiments \cite{xray}. After equilibration, the numerical estimates for  
the diffusion coefficient become at least one order of magnitude smaller than the measured ones.

Prompted by these results and by our own observations on the structure of water around 
iron aqua ions \cite{Sit}, we have undertaken an extensive 
investigation of the static and dynamical properties of water, to ascertain its phase stability around
ambient conditions as predicted by first-principles molecular dynamics.
Particular care has been given to the statistical accuracy of the results, assuring that
the time scales and length scales of the simulations were chosen appropriately for the given conditions.
The paper is organized as follows: 
In section II, we detail all the technical aspects of our simulations.
Section III surveys the static and vibrational properties of the water molecule and 
the water dimer in vacuum, at the GGA-PBE \cite{PBE} density-functional level.
In section IV, we discuss our extensive liquid water simulations, performed with Car-Parrinello 
molecular dynamics, in the temperature range between 325 K and 400 K. Section V discusses the
limitations of this approach, and some of the possible reasons to explain the remaining discrepancies with
experimental results.

\section{Car-Parrinello Molecular Dynamics}

Our first-principles calculations are based on density-functional theory,
periodic-boundary conditions, plane-wave basis sets, and norm-conserving \cite{tm} or 
ultrasoft pseudopotentials \cite{Van}
to represent the ion-electron interactions, as implemented in the public domain codes CP and
PWSCF in the $\nu$-ESPRESSO package \cite{Espresso}.

In Car-Parrinello molecular dynamics,
an extended  Lagrangian is introduced to include explicitly 
the wavefunction degrees of freedom, that are evolved ``on-the-fly''
simultaneously with the ionic degrees of freedom:
\begin{eqnarray}
L_{CP} & = &
\mu\sum_{i}\int d{\bf r} \left|\dot{\Psi}_i\left({\bf r}\right)
\right|^2+\frac{1}{2}\sum_{I}M_I{\bf \dot{R}}_I^2-  \nonumber \\ & &
E_{KS}
\left[\left\{\Psi_i\right\},
{\bf R}_I\right]+ \nonumber \\ & &
\sum_{ij} \Lambda_{ij}\left(\int d{\bf r} \Psi_i^*\left({\bf r}\right)
\Psi_j\left({\bf r}\right)-\delta_{ij}\right).
\label{lagrangian}
\end{eqnarray}
In Eq.~\ref{lagrangian}, $\Psi_i\left(r\right)$ are the occupied Kohn-Sham
orbitals, $\mu$ is a fictitious mass parameter used to control the evolution of
the electronic degrees of freedom in time, $M_I$ are ion masses,
$E_{KS}$ is the Kohn-Sham energy and $\Lambda_{ij}$ are Lagrange
multipliers, used to impose the orthonormality constraint
$\int\Psi_i^*\Psi_j=\delta_{ij}$.  
An extensive review of the method can be found in Ref. \cite{CPref}; 
the subtle technical issues arising in the simulations
of liquid water have been discussed in the recent literature \cite{GrossmanI}$^,$\cite{Schwegler2}. 
We note in passing that our simulations' parameters agree with the
recommendations set forth in these latter papers.
It should be stressed that full convergence of the forces acting on the nuclei with respect to the basis set
is easily achieved with plane waves, and no Pulay forces arise in the dynamics, since the
basis set is independent from the atomic positions.

\subsection{Technical Details}

The structural and vibrational properties of the water monomer and dimer and the binding energy of the
dimer have been calculated using density-functional theory in the generalized-gradient
approximation and the total energy pseudopotential method, and density-functional
perturbation theory \cite{linear}, as implemented
in PWSCF. We performed separate calculations using 
either norm-conserving pseudopotentials for both the hydrogen and the oxygen, or ultrasoft ones.
These same pseudopotentials were also used for the norm-conserving or ultrasoft molecular
dynamics simulations. In particular, the O Troullier-Martins norm-conserving
pseudopotential was generated using the FHI98PP package \cite{fhi98pp} with
core radii for the $s$, $p$ and $d$ components of 1.25 a.u., 1.25 a.u., and 1.4 a.u. respectively. The
Troullier-Martins hydrogen pseudopotential was generated using the Atom code \cite{gian} with a core
radius for the $s$ component of 0.8 a.u. 
The ultrasoft pseudopotentials were taken from the standard
PWSCF distribution \cite{USPP}.
The Kohn-Sham orbitals and charge density have been expanded in plane waves up to a kinetic energy cutoff 
of 25 and 200 Ry (respectively) for the ultrasoft case, and of 80 Ry and 320 Ry for the 
norm-conserving case.  A cubic supercell of side 30 a.u. was used; interaction with periodic images is 
negligible \cite{sprik96} with this unit cell size.

\section{Water monomer and dimer: Structural and Vibrational Properties}

\begin{table}
\caption{Structural properties of the water monomer and dimer and binding energy of the dimer,
as obtained in DFT-PBE using ultrasoft or norm-conserving pseudopotentials, and compared to
available experimental and theoretical results.}
\begin{ruledtabular}
\begin{tabular}{l|l|l|l|l|l}
 &  PBE US & PBE NC & PBE NC            & Expt ~\cite{expt1}$^-$\cite{expt4} & BLYP ~\cite{sprik96}   \\
 & (This   & (This  & (Ref \cite{Schwegler2}&                                    &                     \\
 & work)   & work)  &                   &                     \\
\tableline
$\angle_{HOH}$ & $104.6^{0}$ & $104.2^{0}$ & $104.2^{0}$ & $104.5^{0}$ & $104.4^{0}$ \\
$d_{OH}$(\AA)  &  0.98  & 0.97   & 0.97  & 0.96 & 0.97  \\
$\angle_{OHO}$ & $173^{0}$  & $172^{0}$ & $174^{0}$ & $174^{0}$ & $173^{0}$ \\
$d_{OO}$(\AA)  &  2.89  & 2.88 & 2.90 & 2.98 & 2.95  \\
$E_{dimer}$   &  -23.2 & -23.8 & -21.4 & -22.8 & -18  \\
(kJ/mol)              &        &       &       &     & 
\end{tabular}
\end{ruledtabular}
\label{modim}
\end{table}

The equilibrium structures and energetics are summarized in Table~\ref{modim}. We have included 
published results \cite{sprik96} using the BLYP functional for comparison.
Both ultrasoft and norm-conserving PBE density functionals show very good agreement with experimental
values. In particular, the PBE results have a dimer binding energy in closer agreement to the experiments than
BLYP; the binding energy in this latter case is too weak by 4 kJ/mol, and exhibits a longer O-O distance. 

Table \ref{mfreq} and \ref{dfreq} 
show respectively the vibrational frequencies of the water monomer and dimer in vacuum. In this calculation, a 
hydrogen
mass of 1 a.m.u. was used (as opposed to the 2 a.m.u. mass used for the dynamical simulations of heavy water).
The calculations of the vibrational frequencies with PWSCF using a cubic cell of size (30 a.u.)$^3$. 
To achieve a convergence of a few cm$^{-1}$ in the frequencies, cutoffs of 35 Ryd and 420 Ryd were used 
for wavefunctions
and charge densities, respectively, in the ultrasoft case and, in the norm-conserving case, 100 Ryd and 
400 Ryd were used.
As shown in Table \ref{mfreq} and \ref{dfreq},the PBE functional gives intramolecular stretching modes that 
are in general blue-shifted compared to BLYP and experimental results.
In the calculations for the dimer, the libration modes are also higher for the PBE functionals 
then those given by experiments and BLYP. We note in passing that the errors on these frequencies (especially
the low energy ones) are slightly larger than usually expected from DFT). We will return to this point in later 
section.

\begin{table}
\caption{Vibrational frequencies of water monomer: $\nu_{1}$, $\nu_{2}$ and $\nu_{3}$
are the symmetric stretching, bending and asymmetric modes, respectively.}
\begin{ruledtabular}
\begin{tabular}{l|l|l|l|l}
                    &  PBE (US) & PBE (NC) & Expt ~\cite{expt5} & BLYP ~\cite{sprik96}   \\
\tableline
$\nu_{1}(cm^{-1})$  &  3781  & 3704 & 3657   & 3567  \\
$\nu_{2}(cm^{-1})$  &  1573  & 1599 & 1595   & 1585  \\
$\nu_{3}(cm^{-1})$  &  3908  & 3816 & 3756   & 3663  \\
\end{tabular}
\end{ruledtabular}
\label{mfreq}
\end{table}

\begin{table}
\caption{Vibrational frequencies of water dimer: $\nu_{1}$, $\nu_{2}$ and $\nu_{3}$
are the symmetric stretching, bending and asymmetric stretching modes, respectively. 
Proton acceptor and donor molecules are denoted as (A) and (D). $\nu(Hb)$ are the two 
libration modes between molecules and $\nu(O-O)$ is the hydrogen-bond stretching mode.}
\begin{ruledtabular}
\begin{tabular}{l|l|l|l|l}
                    &  PBE (US) & PBE (NC) & Expt ~\cite{expt1}$^,$~\cite{expt6} & BLYP ~\cite{sprik96}   \\
\tableline
$\nu_{1}(A)(cm^{-1})$  &  3778  & 3695   & 3622   & 3577  \\
$\nu_{2}(A)(cm^{-1})$  &  1570  & 1596   & 1600   & 1593  \\
$\nu_{3}(A)(cm^{-1})$  &  3901  & 3804   & 3714   & 3675  \\
$\nu_{1}(D)(cm^{-1})$  &  3601  & 3532   & 3548   & 3446  \\
$\nu_{2}(D)(cm^{-1})$  &  1593  & 1616   & 1618   & 1616  \\
$\nu_{3}(D)(cm^{-1})$  &  3871  & 3781   & 3698   & 3647  \\
$\nu(Hb)(cm^{-1})$     &  666   & 644    & 520    & 600   \\
$\nu(Hb)(cm^{-1})$     &  379   & 378    & 320    & 333   \\
$\nu(O-O)(cm^{-1})$    &  202   & 196    & 243    & 214   \\
\end{tabular}
\end{ruledtabular}
\label{dfreq}
\end{table}

\section{Liquid water simulations}

\subsection{Liquid water simulation at 325 K}

\subsubsection{Simulation Details}

In this first simulation, we used a body-centered-cubic supercell with 32 heavy water molecules, periodic 
boundary conditions, and the volume corresponding to the experimental \cite{diffexpt} density of
1.0957 $g/cm^3$ at 325 K. 
A body-centered-cubic supercell strikes the optimal balance, for a given volume, in the distance 
between a molecule and its periodic neighbors, and the number of these periodic neighbors.
Ultrasoft pseudopotentials were first used, as detailed in the previous section, 
with plane-wave kinetic energy cutoffs of 25 Ry (wavefunctions) and 200 Ry (charge densities). 
The deuterium mass was used in place of hydrogen to allow for 
a larger time step of integration. 
It should be noted that for classical ions this choice does not affect thermodynamic properties such as the
melting temperature (the momentum integrals for the kinetic energy factor out in the Boltzmann averages). 
Of course, dynamical properties such as the diffusion coefficient will be affected by our choice of
heavier ions. Extensive experimental data for deuterated (heavy) water are in any case widely available.
 
The wavefunction fictitious mass ($\mu$) is chosen to be 700 a.u.; this results in a factor of $\sim$14 between
the average kinetic energy of the ions and that of the electrons.
A time step of 10 a.u. was used to integrate the electron and ionic equations of motions. 
This combined choice of parameters allows for roughly 25 ps of simulation time
without a significant drift in the kinetic energy of wavefunctions 
and the constant of motion for the Lagrangian (\ref{lagrangian}). 
Our choice of fictitious mass is consistent with the ratio
$\mu/M\le\frac{1}{3}$ for heavy water molecules suggested by Grossman et al.\cite{GrossmanI} , and assures that
the physical properties are not influenced by the electronic degrees of freedom. 
Our initial configuration was obtained from
a comparatively short 1.2 ps simulation at twice the value of the target temperature (650 K instead of 325 K); a restart
with zero initial velocities was then performed at 325 K, with the temperature controlled by a single
Nose-Hoover thermostat on the ions (no electronic thermostat is applied in any of the simulations).

\subsubsection{Results}

\begin{figure}
\centerline{
\rotatebox{-90}{\resizebox{2.8in}{!}{\includegraphics{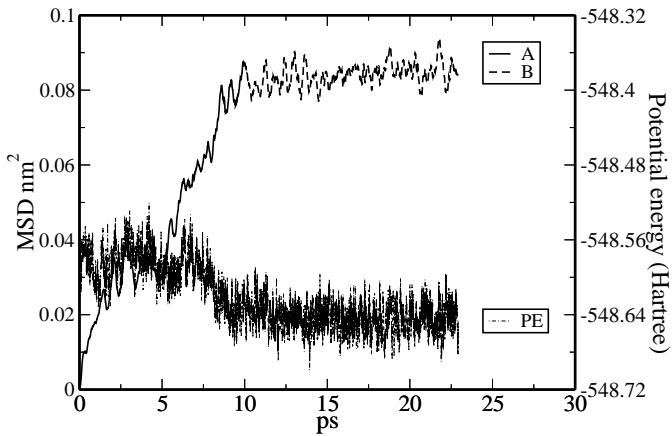}}}
}
\caption{Mean square displacement and potential energy as a function of time for our first 32-water 
molecules simulations at 325 K.
A: Diffusive region in the first stage of the simulation.
B: After about 10 ps diffusivity drops abruptly.}
\label{therm}
\end{figure}

\begin{figure}
\centerline{
\rotatebox{-90}{\resizebox{2.8in}{!}{\includegraphics{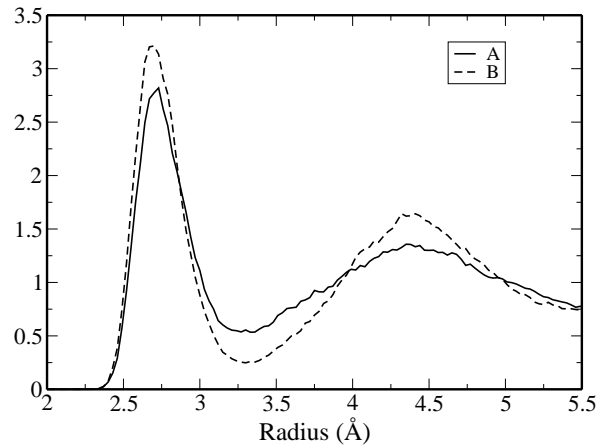}}}
}
\caption{O-O radial distribution function before (A) and after (B) the equilibrium is attained.}
\label{grth}
\end{figure}

The thermostat stabilizes quickly ($\sim$1 ps) the temperature around the target value of 325 K, with the usual 
fluctuations due to small size of the system. 
However, the system is still far from equilibrium; this can be clearly observed by looking at the time
evolution of the radial distribution function (RDF) and mean square displacements (MSDs).
We plot in Fig.~\ref{therm} the MSDs of the oxygen atoms as a function of time. 
For the first 10 ps, the water molecules diffuse with a velocity comparable with
experimental data; after 10 ps, a sharp drop in
diffusivity is observed, accompanied by a distinctive sharpening of the features in 
the oxygen-oxygen radial distribution function (see Fig.~\ref{grth}). 
The radial distribution
functions was calculated from the infinite bulk system by repeating
the unit cell in all directions. This means that molecules up to
5.5 \AA\ in a 32-molecule cell are inequivalent.
As we move beyond this cutoff distance,
the radial distribution functions will include both molecules that are inequivalent,
and some that are equivalent. The statistical accuracy is going to be gradually, but slowly,
affected. In this graph, and most of the latter figures, we plot radial distribution functions
only up to 5.5 \AA (but see e.g. Fig. 10 for a discussion of the finite-size effects).

The potential energy
for the dynamics (Fig.~\ref{therm}) also drifts downward in these first 10 ps, and
stabilizes afterward.
We calculated the self-diffusion coefficient ($D_{self}$) from the Einstein relation
(in 3 dimensions):
\begin{equation}
6D_{self} = \lim_{t\rightarrow\infty}\frac{d}{dt}
\left<\left|{\bf r}_i\left(t\right)
-{\bf r}_i\left(0\right)\right|^2\right>.
\label{diffcoef}
\end{equation}
The structural and dynamical properties before and after this 10 ps mark are summarized in Table~\ref{diffth};
experimental values at 298 K are included for comparison.
All these observation conjure to a picture in which the system takes at least
10 ps to reach a reasonably thermalized state, in a process somewhat reminiscent of a glass
transition.
Although the time needed for equilibration will be dependent on the initial conditions,
these preliminary result suggests that simulation times in the order of ten of picoseconds might
be need to calculate well-converged thermodynamic observables.

Once the initial thermalization trajectory was discarded from our averages, we obtained a
self-diffusion coefficient one order of magnitude smaller than the experimental value measured at room temperature. 
This result, combined with the clear over-structuring of the oxygen-oxygen radial distribution function $g_{OO}(r)$, 
indicates that our system has reached a ``frozen'' equilibrium state very different from what expected
for liquid water (strictly speaking, a system with a finite and small number of
inequivalent atoms or molecules will never undergo a phase transition).

These considerations indicates the need to assess accurately the phase stability of liquid water
as obtained from first-principle molecular dynamics. 
At the same time, they point out the requirement of
long simulation times, and a careful analysis of the technical details of the simulations.

\begin{table}
\caption{Structural and dynamical parameters before and after the 10 ps mark, compared with the
experimental results at 298 K.}
\begin{ruledtabular}
\begin{tabular}{l|l|l}
   & $g(r)_{max}$ & D $(cm^{2}/s)$ \\
\tableline
 Before & 2.82 & $1.5 \times 10^{-5}$  \\
 After  & 3.21 & $0.14 \times 10^{-5}$  \\
 Expt \cite{diffexpt}$^,$\cite{neutron} & 2.75 & $2.0 \times 10^{-5}$ \\
\end{tabular}
\end{ruledtabular}
\label{diffth}
\end{table}

\subsection{Extensive water simulations in the region between 325 K and 400 K}

\subsubsection{Simulation details}

\begin{figure}
\centerline{
\rotatebox{-90}{\resizebox{2.8in}{!}{\includegraphics{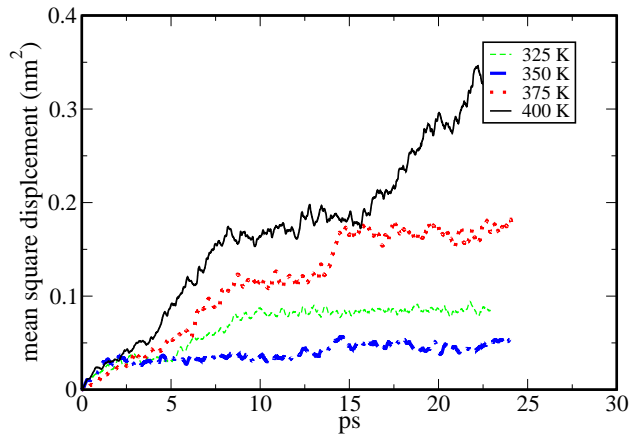}}}
}
\caption{Mean square displacements as a function of time of our initial equilibration runs for 32 water molecules at 325, 350, 375 and 400 K.}
\label{flat}
\end{figure}

In order to find out the temperature at which our system would move from a glassy to a liquid-like
state, we decided to perform a series of extensive simulations at increasing temperature, from 325 K to
400 K.
We first performed $\approx 25$ ps simulations at 325 K, 350 K, 375 K and 400 K, using $\mu$=700 a.u. and 
$\delta t$=10 a.u. . We used at every temperature the experimental densities \cite{diffexpt}, with the caveat that
the 400 K value was obtained by extrapolation. 25 ps is approximately the maximum time allowed for a
simulation with these parameters before the drift in the kinetic energy of the wavefunctions becomes
apparent. We thus used these simulations as efficient ``thermalization'' runs,
to be followed by production runs that will be described below. It is interesting to monitor during these
thermalization runs the evolution of the MSDs; these are shown for all four temperatures in 
Fig.~\ref{flat}. An abrupt drop in diffusivity is observed for all cases
but one (predictably, the one at the highest temperature of 400 K).
The onset of this drop in diffusivity varies, but broadly speaking is again of the order of 10 ps.

\begin{figure}
\centerline{
\rotatebox{-90}{\resizebox{2.9in}{!}{\includegraphics{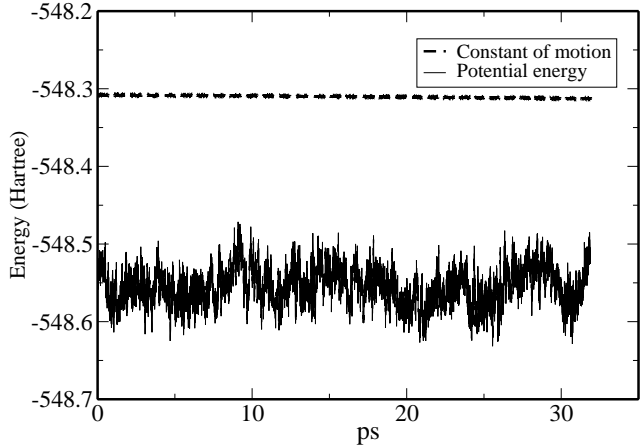}}}
}
\caption{Potential energy and constant of motion in a production run at 400 K.}
\label{400con}
\end{figure}

\begin{figure}
\centerline{
\rotatebox{-90}{\resizebox{2.9in}{!}{\includegraphics{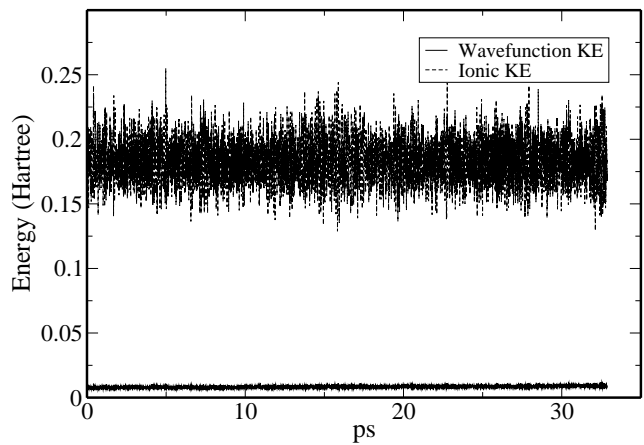}}}
}
\caption{Kinetic energy of the ions and the electrons in a production run at 400 K.}
\label{400KE}
\end{figure}

With these trajectories, now well thermalized in configuration space, we started our
four production runs at 325 K, 350 K, 375 K and 400 K, and each of them starting 
from the last ionic configurations of the previous simulation at the corresponding temperature, but  with 
zero ionic velocities. These four runs, lasting between 
20 ps and 37 ps, were performed using $\mu$=450 a.u. and $\delta t$=7 a.u.
This choice of mass and timestep allows for an excellent conservation of the constant of motion, and negligible
drift in the fictitious kinetic energy of the electrons, for the simulation times considered. 
The ratio between the kinetic energy of the ions and that of the electrons was $\approx$ 22 for the 
whole production time. Although a small enough fictitious mass decouples the
electronic and ionic degrees of freedom, Tangney et. al.~\cite{Tangney} pointed out that there 
is a fictitious mass dependent error that is not averaged in the time scale of ionic motions.
Schwegler et. al.~\cite{Schwegler2} studied this effect comparing closely Car-Parrinello and Born-Oppenheimer 
simulations finding a larger self-diffusion coefficient in the Car-Parrinello
simulation. However, the structural and thermodynamical properties were not affected.

\begin{figure}
\centerline{
\rotatebox{-90}{\resizebox{2.9in}{!}{\includegraphics{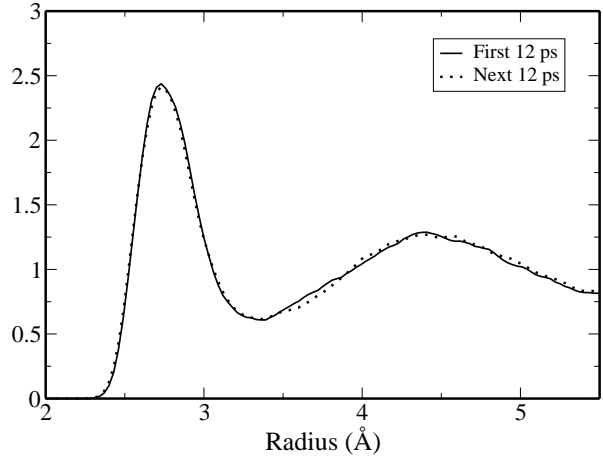}}}
}
\caption{O-O radial distribution functions calculated from the first and the next 12 ps of the simulation
at 400 K.}
\label{400gr}
\end{figure}

We show in Figs.~\ref{400con} and ~\ref{400KE} the case of the 400 K simulation; we stress that no periodic quenching 
of the electrons was needed, and the simulations were single uninterrupted runs.
Since the initial configurations were already at equilibrium at their respective temperature, and thermalization
in momentum space is fast, we found that ``production time''
can start early in the simulations. We discarded from each trajectory the initial 1.2 ps that were needed to
allow the ions to reach their target kinetic energy. As a measure of the good thermalization reached
in the simulations, we show in Fig.~\ref{400gr} the O-O radial distribution function obtained from the
first 12 ps of our 400 K trajectory, and the following 12 ps.

To rule out any spurious effect in our simulations coming from the use of pseudopotentials, or an
extended Lagrangian, we also performed two simulations
using norm-conserving pseudopotentials (as described in Section II). These require
larger plane-wave basis sets (80 Ry for the wavefunctions and 320 Ry charge densities, corresponding to 40000 plane
waves vs 7000 for the ultrasoft case), and discrepancies, if any, with the
ultrasoft calculation will provide an approximate estimate of the effects of the pseudopotential approximation and 
of the dynamics of the fictitious degrees of freedom.
For the runs involving these norm-conserving pseudopotentials, we used $\mu=$300 a.u. and $\delta 
t$=5 a.u\cite{timestep1}.
These parameters results in a factor of $\sim$ 13 between the kinetic energy of the ions and that of the electrons.
Details of all these simulations are shown in Table~\ref{details}. 

\begin{table}
\caption{Details of the production runs.}
\begin{ruledtabular}
\begin{tabular}{c|c|c|c|c|c}
Pseudo- & T(K) & Density  & $\mu$ & $\delta t$ & Production  \\
potentials & & ($g/cm{^3}$) & (a.u.) & (a.u.) & time (ps)\\
\tableline
US &325    & 1.0957 & 450 & 7 & 37.6\\
NC &325    & 1.0957 & 300 & 5 & 22.1\\
US &350    & 1.0815 & 450 & 7 & 22.9\\
US &375    & 1.0635 & 450 & 7 & 21.1\\
US &400    & 1.0554 & 450 & 7 & 32.5\\
NC &400    & 1.0554 & 300 & 5 & 20.2\\
\end{tabular}
\end{ruledtabular}
\label{details}
\end{table}

\subsubsection{Results}

\begin{figure}
\centerline{
\rotatebox{-90}{\resizebox{2.9in}{!}{\includegraphics{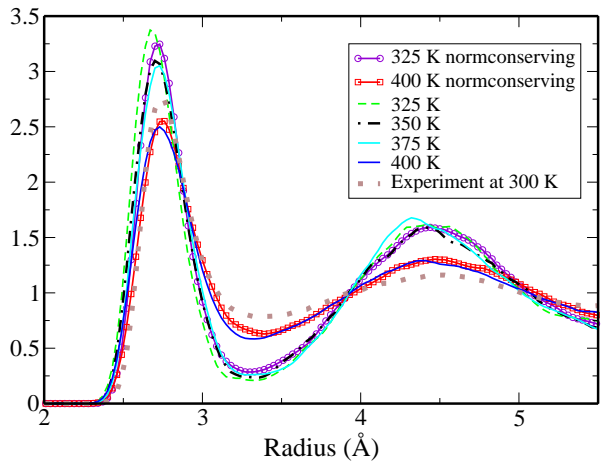}}}
}
\caption{O-O radial distribution functions calculated from the production runs at 325 K, 350 K, 375 K and 400 K 
for ultrasoft and norm-conserving pseudopotentials. Experimental result is taken from Ref \cite{neutron}.}
\label{grdT}
\end{figure}

\begin{figure}
\centerline{
\rotatebox{-90}{\resizebox{2.9in}{!}{\includegraphics{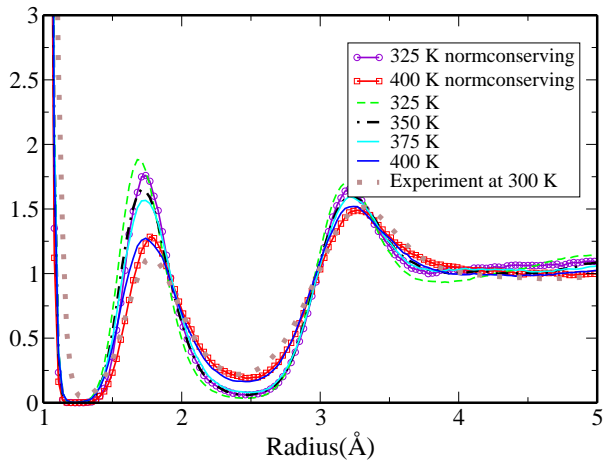}}}
}
\caption{O-D radial distribution functions calculated from simulations at 325 K, 350 K, 375 K and 400 K
for ultrasoft and norm-conserving pseudopotentials.
Experimental result is taken from Ref \cite{neutron}.}
\label{groh}
\end{figure}

\begin{figure}
\centerline{
\rotatebox{-90}{\resizebox{2.9in}{!}{\includegraphics{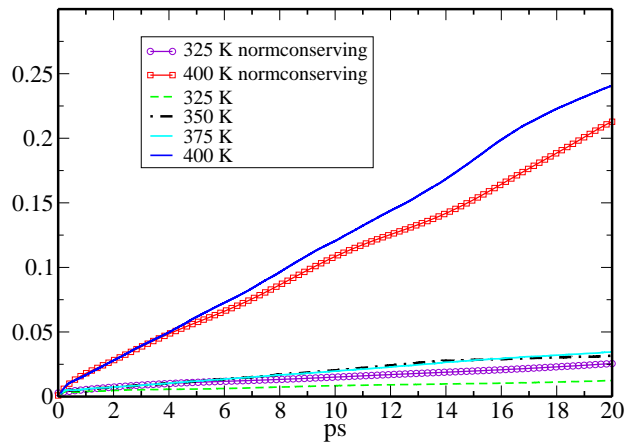}}}
}
\caption{Mean square displacements calculated from simulations at 325 K, 350 K, 375 K and 400 K
for ultrasoft and norm-conserving pseudopotentials.}
\label{diffdT}
\end{figure}

The oxygen-oxygen and oxygen-deuterium
radial distribution functions for the different conditions considered in this work are shown in
Fig.~\ref{grdT} and Fig.~\ref{groh} respectively.
We find that at temperatures of 375 K and below both $g_{OO}$ and $g_{OH}$ show considerably more structure
than found experimentally at 300 K. For this range of temperatures the height of the first peak 
of $g_{OO}(r)$ is roughly between 3.2 and 3.4, and significantly larger than the experimental value of 2.75
(also measured at 300 K). 
However, when the temperature in our simulations is increased to 400 K, a distinct drop of the first peak to
$\sim$ 2.5 is observed, the radial distribution function $g_{OO}(r)$ and $g_{OH}(r)$ show a sharp change in their
structure, and the water molecules start diffusing much faster, as reflected in
the MSDs curves for the oxygen atoms shown in Fig.~\ref{diffdT}.
To provide cleaner statistics, the MSDs curves shown have been calculated as an average over 
individual MSDs curves, each obtained from our trajectory by shifting - for each individual MSDs curve - the
starting configuration by 0.017 ps (in this way, a 17 ps trajectory would provide 1000 progressively shorter
MSDs curves that are then averaged). The
self-diffusion coefficient $D_{self}$ is calculated from 
the slope of the respective MSDs curve in the range of 1 to 20 ps using Einstein's relation (\ref{diffcoef}).
Negligible differences are observed between simulations performed with 
ultrasoft or norm-conserving pseudopotentials, ruling out any role of the pseudopotential details in 
this observed result.

The structural and dynamical results are summarized in Table~\ref{dT}.
As seen in this table, there is an eight-fold increase in $D_{self}$ when increasing the temperature from
375 K to 400 K.
Price et. al.\cite{Price} reported the experimental self-diffusion coefficient of supercooled heavy water at different
of temperatures. At 276.4 K, which is just below the freezing temperature of heavy water (277.0 K), the experimental 
value for $D_{self}$ is 
$0.902\times 10^{-5} cm^2/s$. $D_{self}$ for our simulations at 325 K, 350 K and 375 K is
0.16, 0.25 and 0.26 $\times 10^{-5} cm^2/s$, respectively.
These numbers are significantly smaller then the experimental value below the freezing point. On the other hand,
$D_{self}$ at 400 K in our simulations is comparable to the experimental value at 300 K.
These observations suggest that the theoretical freezing point for water at the DFT-PBE level is between 375 K and 400 K,
and water below 375 K is in a glassy/supercooled state.

The hydrogen-bond structure can be studied calculating the number of hydrogen bonds per molecule:
we identify a hydrogen bond is identified when two oxygen atoms are closer than 3.5 \AA$\:$and the
$\angle_{OHO}$ angle is greater than 140$^{o}$ (consistently with Ref \cite{Hbddef1}, and at slight variance with Ref \cite{Hbddef2}). 
The results are shown in the last column of Table~\ref{dT}.
Between the temperatures of 325 K and 375 K there are only small changes in the number of hydrogen bonds in
the system, going from 3.86 per molecule to 3.72 per molecule.
An abrupt decrease to 3.37-3.45 bonds occurs when the temperature
increases from 375 K to 400 K. The experimental value of 3.58 at 300 K lies between our values 
at 375 K and 400 K, further suggesting that the freezing point in our simulations is between 375 K and 400 K.

In summary, we found clear liquid-like signatures in the structure and dynamics of
(heavy)  water, as described by
DFT-PBE and Car-Parrinello MD, for temperatures reaching at least 400 K. 
At temperatures of 375 K and lower water is found to be in a glassy state, more structured and with much lower 
diffusivity. This discrepancy of more than 100 K between experimental and theoretical results is
obviously relevant, given the enormous importance of water in the description of systems ranging
from electrochemistry to biology, and it is investigated further in the next section.

\begin{table}
\caption{Summary of structural and dynamical properties of water. $D_{self}$ is self-diffusion coefficient.
$g_{max}$ is first peak height and R[$g_{max}$] is location of first peak. The last column is the number
of Hydrogen-bond per molecule.}
\begin{ruledtabular}
\begin{tabular}{c|c|c|c|c}
            & $D_{self}$  & $g_{max}$ &R[$g_{max}$] & no.of H-bonds \\
            & $10^{-5}cm^2/s$   &      &         & per molecule   \\ \hline
  325 K (US) & 0.07 & 3.38 & 2.67 & 3.86\\
  325 K (NC) & 0.16 & 3.25 & 2.73 & 3.79\\
  350 K (US) & 0.25 & 3.25 & 2.73 & 3.77\\
  375 K (US) & 0.26 & 3.10 & 2.71 & 3.72\\
  400 K (US) & 2.03 & 2.50 & 2.73 & 3.45\\ 
  400 K (NC) & 1.66 & 2.55 & 2.75 & 3.37\\
Expt\cite{diffexpt}$^,$\cite{hbexpt} at 300 K & 1.80 & 2.75 & 2.80 & 3.58 \\ 
\end{tabular}
\end{ruledtabular}
\label{dT}
\end{table}

\section{Overestimation of the freezing temperature of water}

There are several possible reasons for the overestimation of the freezing temperature of water, and 
several of them could play a significant role. We discuss here some of the possibilities.

\subsection{Finite-size effects}

\begin{figure}
\centerline{
\rotatebox{0}{\resizebox{3.8in}{!}{\includegraphics{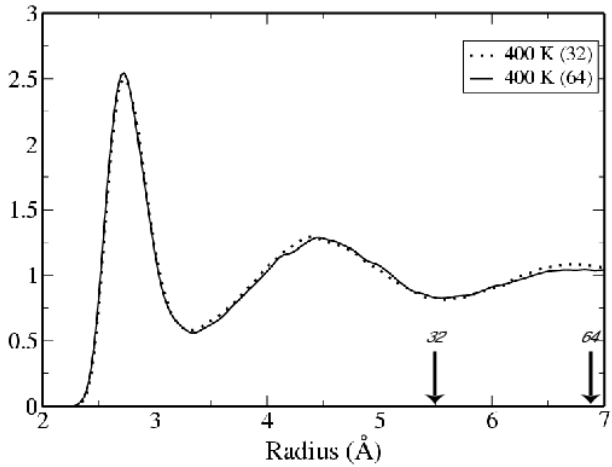}}}
}
\caption{O-O radial distribution function for a Car-Parrinello simulations with 32 or 64 molecules. 
The arrows indicate the radius of the sphere that is completely
inscribed by the BCC simulation cells with 32 or 64 molecules.}
\label{64_32}
\end{figure}

Since our simulation cell contains only 32 water molecules, finite-size effects could obviously play a role even
if periodic-boundary conditions are used. The interactions of water molecules with their periodic 
images could be considerable due to the long-range hydrogen-bond network. On the other hand, when
zero correlations are found between a molecule and its periodic image we can safely assume that the unit
cell is for all practical purposes large enough, and every molecule feels the same environment that
it would have in an infinite system. In our case, the distance between a molecule and its
eight periodic images is $\sim$ 11 \AA\, and at this distance all radial
distribution functions look very flat and unstructured. In any case, to 
study the finite-size effects 
we carried out another extensive simulation (40 ps total, with 15 ps of production time 
following 25 ps of thermalization) for a system composed of 64 heavy water molecules at 
400 K. We used the same parameters for this simulation
as in the 32-molecule, 400 K ultrasoft simulation. The oxygen-oxygen radial distribution
function $g_{OO}$(r) is shown in Fig.~\ref{64_32} for both the 32- and 64-molecule systems.
As mentioned previously, the radial distribution
functions was calculated by repeating
the unit cell in all directions.  The molecules up to
5.5 \AA\ in the 32-molecule cell, and 6.9 \AA\ in the 64-molecule cell are inequivalent.
We indicate in the graphs with two arrows, the radii of the spheres completely
inscribed by in BCC simulation cells.

\begin{figure}
\centerline{
\rotatebox{-90}{\resizebox{2.9in}{!}{\includegraphics{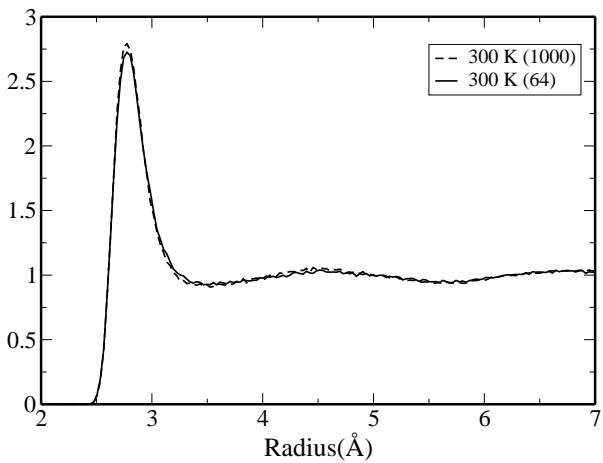}}}
}
\caption{O-O radial distribution function for a classical (SPC) simulation with 64 or 1000 water 
molecules.}
\label{classical}
\end{figure}

Differences between 32- and 64-molecule systems 
are negligible, and within the variance for simulations of the order of 10-20 ps
(as estimated from uncorrelated classical simulation data\cite{GrossmanI});
the 64-water simulation shows a marginally more structured $g(r)$ where the first peak height is at about 
2.6 (compared to 2.5 for the 32-water case).
Larger ab-initio simulations would be too demanding; for this reason, we 
performed two classical simulations at 300 K using the SPC force field \cite{SPC} for water, and
comparing the case of 64 and 1000 water molecules (each simulation lasting 1000 ps). The $g_{OO}$(r) calculated 
from these two runs are shown in Fig.~\ref{classical}, and again we do not find
any significant differences between these two curves. These results help ruling out finite-size effects
as the major cause of the discrepancy observed with the experimental numbers.

\subsection{Exchange-correlation functional effects}

While density-functional theory is in principle exact, any
practical application requires an approximated guess to the true
exchange-correlation functional. In this work, we have used the GGA-PBE
approach \cite{PBE}. As it was observed in Sec. III, the structural
properties for the water molecule and dimer are in excellent agreement
with experiments, as is the binding energy for the dimer. On the other hand,
the vibrational properties show larger discrepancies with experiments than
usually expected, in particular for some of the libration modes in the dimer.
This result certainly points to the need for improved functionals to describe
hydrogen bonding. The dependence of the melting point on the
exchange-correlation functional chosen is more subtle; below 400 K, PBE water
displays solid-like oxygen-oxygen radial distribution functions that are
only slightly affected by the temperature, and that are similar to those
obtained with a fairly different functional such as BLYP \cite{GrossmanI}. 
The similarity between these radial distribution functions is
just a reminder that the structural property and the geometry
of the intermolecular bonds are well described by different functionals;
once water is ``frozen'', all radial distribution functions will look similar.
The temperature at which this transition takes place could be affected by
the use of different functionals \cite{Vande} and the magnitude of the contribution of
any one of them to the melting point temperature is still open to
investigation.

\subsection{Quantum effects}

\begin{figure}
\centerline{
\rotatebox{-90}{\resizebox{2.9in}{!}{\includegraphics{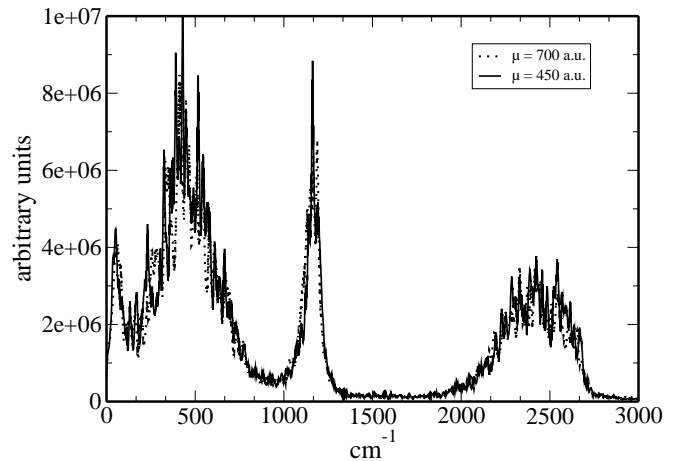}}}
}
\caption{Power spectrum of deuterium atoms calculated from the velocity-velocity correlation function.
For comparison, we also show the power spectrum as obtained with a larger fictitious mass of 700 a.u 
(dotted line) instead of 450 a.u.}
\label{Hpowsp}
\end{figure}

In first-principles Car-Parrinello or Born-Oppenheimer molecular dynamics simulations the ions are most
often treated as classical particles, which is a good approximation for heavy ions (path-integral simulations
can describe the quantum nature of the ions \cite{PIMD1}$^-$\cite{PIMD3}, but their computational costs, when
paired with a first-principles DFT descriptions of the electrons, preclude at this moment simulations
with the statistical accuracy needed).
However, for light ions like hydrogen or deuterium, the effects of a proper quantum statistics
can be very significant; tunneling of the nuclei can also affect the dynamics 
\cite{Quantumeff1}$^,$\cite{Quantumeff2}. 
In the case of water, all the intramolecular vibrational modes and some of the intermolecular modes
are much higher in energy compared to room temperature. 
We show in Fig.~\ref{Hpowsp} the power spectrum for the deuterium atoms as calculated from the velocity-velocity
correlation function of heavy water molecules for our simulations at 400 K (ultrasoft, 32 molecules). 
The distinctive peaks of the intramolecular stretching
and bending modes are centered around 2400 cm$^{-1}$ and 1200 cm$^{-1}$ respectively, much
higher than the room temperature of $k_{B}T_{room} \approx$ 200 cm$^{-1}$. The peak at 500 cm$^{-1}$ corresponds 
to the intermolecular vibrational modes, also larger than
$k_{B}T_{room}$.
When ions are treated as classical particles, as in our ab-initio molecular dynamics simulations, all 
vibrational modes obey Boltzmann statistics. In reality,
modes with frequency higher than $k_{B}T_{room}$ 
are frozen in their zero-point motion state, and their exchange of energy 
with the lower-frequency modes (``the environment'') is suppressed - in other words their contribution
to the specific heat is zero, in full analogy with the low-temperature discrepancies from the
Dulong-Petit law in solids. This effect could significantly affect
the dynamics of water-water interactions, and it has long been argued that treating each
water molecule as a rigid body could actually provide a closer match with experimental 
conditions. In fact,
recent ab-initio simulations \cite{Allesch} in which the water molecules are constrained to
maintain their equilibrium intramolecular bond lengths and bond angle result in a more diffusive
and less structured description of water, that remains liquid at a temperature of 326 K.
Path-integral simulation \cite{Rossky} for water described with classical potentials
did find as well a significant difference due to the quantum effects
(i.e. the freezing of the high-energy vibrational excitation), of
the order of 50K.

\section{Conclusion}

We performed extensive first-principles molecular dynamics simulations of heavy water at the DFT-PBE
level. Equilibration times
are found to be comparatively long, and easily in excess of $\approx$ 10 ps at ambient temperature. 
Good statistics was obtained with simulations that included at least 25 ps of thermalization time,
followed by 20 to 40 ps of production time.
At ambient temperature, water is found to be over-structured compared to experiment, and with a
diffusivity that is one order of magnitude smaller than expected.
An abrupt change in the structure and dynamics is observed when the the temperature is
raised from 375 K to 400 K; even at this high temperature, where a liquid-like state is
reached, water shows more structure and less diffusivity than found experimentally at 
room temperature. 
These results are broadly independent of some of the possible errors or inaccuracies involved in
first-principles simulations, including in this case finite-size effects, insufficient thermalization or
simulation times, or poorly designed pseudopotentials.
Our simulations suggest that the freezing point is around 400 K - this discrepancy is at variance with
the good agreement for the structure and energetics of the water monomer and dimer with the experimental values,
and could originate in the neglect of quantum statistics for the many high-frequency vibrational modes
in this system. 

This work was performed under the support of 
the Croucher Foundation and Muri Grant DAAD 19-03-1-0169.
Calculations in this work have been done using the $\nu$-ESPRESSO package \cite{Espresso}


\begin{thebibliography}{10}

\bibitem{water}
\textit{Water: A comprehensive Treatise}, edited by F. Franks (Plenum, New York, 1972) Vol 1.

\bibitem{xray}
J.~M. Sorenson, G. Hura, R.~M. Glaeser, and T. Head-Gordon, J. Chem. Phys. {\bf
  113},  9149  (2000).

\bibitem{neutron}
A.~K. Soper, Chem. Phys. {\bf 258},  121  (2000).

\bibitem{emp1}
A. Rahman and F. ~H. Stillinger, J. Chem. Phys. {\bf 55}, 3336 (1971).

\bibitem{emp2}
W. ~L. Jorgensen, J. Chandrasekhar, J. ~D. Madura, R. ~W. Impey, and M. ~L. Klein
J. Chem. Phys. {\bf 79}, 926 (1983).

\bibitem{emp3}
K. Toukan and A. Rahman, Phys. Rev. B {\bf 31}, 2643 (1985).

\bibitem{emp4}
S. W. Rick, S. J. Stuart, and B. J. Berne, J. Chem. Phys. {\bf 101}, 6141 (1994).

\bibitem{emp5}
G. Lamoureux, A. ~D. MacKerell, Jr., and B. Roux, J. Chem. Phys. {\bf 119}, 5185 (2003).

\bibitem{empend}
S. Izvekov, M. Parrinello, C. ~J. Burnham and A. Voth, J. Chem. Phys. {\bf 120}, 10896 (2004).

\bibitem{CPwat1}
K. Laasonen, M. Sprik, and M. Parrinello, J. Chem. Phys. {\bf 99}, 9080 (1993).

\bibitem{CPwat2}
M. Bernasconi, P. L. Silvestrelli, and M. Parrinello, Phys. Rev. Lett. {\bf 81}, 1235 (1998).

\bibitem{CPwat3}
P. ~L. Silvestrelli and M. Parrinello, Phys. Rev. Lett. {\bf 82}, 3308 (1999).

\bibitem{CPwat4}
P. ~L. Silvestrelli and M. Parrinello, J. Chem. Phys. 111, 3572 (1999). 

\bibitem{CPwat5}
E. Schwegler, G. Galli, and F. Gygi, Phys. Rev. Lett. {\bf 84}, 2429 (2000).

\bibitem{CPwat6}
M. Boero, K. Terakura, T. Ikeshoji, C. ~C. Liew, and M. Parrinello, J. Chem. Phys. {\bf 115}, 2219 (2001). 

\bibitem{CPwat7}
S. Izvekov and G. A. Voth, J. Chem. Phys. {\bf 116}, 10372 (2002).

\bibitem{CPwatend}
I-Feng ~W. Kuo and C. ~J. Mundy, Science {\bf 303} 658 (2004).  

\bibitem{CP1}
R. Car and M. Parrinello, Phys. Rev. Lett. {\bf 55}, 2471 (1985).

\bibitem{CP2}
K. Laasonen, A. Pasquarello, R. Car, C. Lee, and D. Vanderbilt, Phys. Rev. B {\bf 47}, 10142 (1993) 

\bibitem{GrossmanI}
J. Grossman, E. Schwegler, E. Draeger, F. Gygi and G. Galli, J. Chem. Phys. {\bf 120}, 300 (2004).

\bibitem{kress}
D. Asthagiri, L. R. Pratt, and J. D. Kress, Phys. Rev. E {\bf 68}, 041505 (2003).

\bibitem{Sit}
P. H.-L.Sit, M. Cococcioni, and N. Marzari, in preparation

\bibitem{PBE}
J. P. Perdew, K. Burke, and M. Ernzerhof, Phys. Rev. Lett. {\bf 77}, 3865 (1996).

\bibitem{tm}
N. Troullier and J. Martins, Phys. Rev. B {\bf 43}, 1993 (1991).

\bibitem{Van}
D. Vanderbilt, Phys. Rev. B 41, 7892 (1990).

\bibitem{Espresso}
S. Baroni, A. Dal Corso, S. de Gironcoli, P. Giannozzi, C.
Cavazzoni, G. Ballabio, S. Scandolo, G. Chiarotti, P. Focher, A.
Pasquarello, K. Laasonen, A. Trave, R. Car, N. Marzari, A. Kokalj,
{\tt http://www.pwscf.org/}.

\bibitem{CPref}
D. Marx and J. Hutter ``Ab-initio Molecular Dynamics: Theory and Implementation'',
Modern Methods and Algorithms in Quantum Chemistry Forschungzentrum Juelich, NIC Series, vol. 1, (2000).
http://www.fz-juelich.de/nic-series/Volume1/marx.pdf

\bibitem{Schwegler2}
E. Schwegler, J. Grossman, F. Gygi and G. Galli, J. Chem. Phys. {\bf 121}, 5400 (2004).

\bibitem{linear}
S. Baroni, P. Giannozzi, and A. Testa, Phys. Rev. Lett. {\bf 59}, 2662-2665 (1987).

\bibitem{fhi98pp}
M.Fuchs and M. Scheffler, Comput. Phys. Commun. {\bf 119}, 67 (1999).
Web site: {\tt http://www.fhi-berlin.mpg.de/th/fhi98md/fhi98PP}.

\bibitem{gian}
Web site: {\tt http://www.nest.sns.it/\~{}giannozz}

\bibitem{USPP} H.pbe-rrkjus.UPF and O.pbe-rrkjus.UPF from the v1.3 pseudopotential table at
{\tt http://www.pwscf.org/pseudo.htm}.

\bibitem{sprik96}
M. Sprik, J. Hutter and M. Parrinello, J. Chem. Phys. {\bf 105}, 1142 (1996).

\bibitem{expt1}
K. Kuchitsu, Y. Morino, Bull. Chem. Soc. Jpn. {\bf 38}, 805 (1965).

\bibitem{expt2}
T. R. Dyke, K. M. Mack and J. S. Muentner, J. Chem. Phys. {\bf 66}, 498 (1997).

\bibitem{expt3}
L. A. Curtiss, D. L. Frurip, and M. J. Blander, J. Chem. Phys. {\bf 71}, 2703 (1979).

\bibitem{expt4}
R. M. Bentwood, A. J. Barnes, and W. J. Orville Thomas, J. Mol. Spectrosc. {\bf 84}, 391 (1980).

\bibitem{expt5}
K. Kuchitsu, Y. Morino, Bull. Chem. Soc. Jpn. {\bf 38}, 805 (1965).

\bibitem{expt6}
R. M. Bentwood, A. J. Barnes, and W. J. Orville Thomas, J. Mol. Spectrosc. {\bf 84}, 391 (1980).

\bibitem{diffexpt}
D. ~J. Wilber, T. Defries, and J. Jonas, J. Chem. Phys. {\bf 65}, 1783 (1976).

\bibitem{Tangney}
P. Tangney and S. Scandolo, J. Chem. Phys. {\bf 116}, 14 (2002)

\bibitem{timestep1}
The use of norm-conserving pseudopotentials requires a larger plane-wave energy cutoff and thus a larger basis set.
A smaller fictitious mass and a small time step are thus required. For extensive discussion, see the above reference.

\bibitem{Price}
W. S. Price, H. Ide, Y. Arata and O. S$\ddot{o}$derman, J. Phys. Chem. B {\bf 104}, 5874 (2000).

\bibitem{Hbddef1}
E. Schwegler, G. Galli, and F. Gygi Phys. Rev. Lett. {\bf 84}, 2429 (2000)

\bibitem{Hbddef2}
M. V. Fern$\acute{a}$ndez-Serra and E. Artacho cond-mat/0407237.

\bibitem{hbexpt}
A. K. Soper, F. Bruni, and M. A. Ricci, J. Chem. Phys. 106, 247 (1997).

\bibitem{SPC}
H. Berendsen, J. Postma, W. van Gunsteren, J. Hermans, In Intermolecular Forces; Pullmann, B., Ed.; Reidel:
Dordrecht, 1981; p 331.

\bibitem{Vande}
J. VandeVondele, F. Mohamed, M. Krack, J. Hutter, M. Sprik and M. Parrinello, J. Chem. Phys. {\bf 122}, 014515 (2005).

\bibitem{PIMD1}
D. Marx and M. Parrinello Z. Phys. B (Rapid Note) {\bf 95}, 143 (1994).

\bibitem{PIMD2}
D. Marx and M. Parrinello, J. Chem. Phys. {\bf 104}, 4077 (1996).

\bibitem{PIMD3}
M. E. Tuckerman, D. Marx, M. L. Klein and M. Parrinello, J. Chem. Phys. {\bf 104}, 5579 (1996).

\bibitem{Quantumeff1}
D. Marx, M. E. Tuckerman, J. Hutter and M. Parrinello, Nature {\bf 397}, 601 (1999).

\bibitem{Quantumeff2}
M. E. Tuckerman, D. Marx and M. Parrinello, Nature {\bf 417}, 925 (2002).

\bibitem{Allesch}
M. Allesch, E. Schwegler, F. Gygi and G. Galli, J. Chem. Phys. {\bf 120}, 5192 (2004). 

\bibitem{Rossky}
R. A. Kuharski and P. J. Rossky, J. Chem. Phys. {\bf 82}, 5164 (1985)

\end{thebibliography}
\end{document}